\documentclass[english,journal=jcpafh]{achemso}
\usepackage[T1]{fontenc}
\usepackage[latin9]{inputenc}
\usepackage{color}
\usepackage{amstext}
\usepackage{graphicx}
\usepackage{subscript}

\makeatletter

\title{Why does B\textsubscript{12}H\textsubscript{12}-icosahedron need
two electrons to be stable: A first-principles electron-correlated
investigation of B\textsubscript{12}H\textsubscript{n}(n=6,12) clusters}

\author{Pritam Bhattacharyya}

\affiliation{Department of Physics, Indian Institute of Technology Bombay, Powai,
Mumbai 400076, India}

\email{pritambhattacharyya01@gmail.com}

\author{Ihsan Boustani}

\affiliation{Theoretical and Computational Chemistry, Faculty of Mathematics and
Natural Sciences, Bergische Universit\"at, Wuppertal, Gauss Strasse
20, D-42097 Wuppertal, Germany}

\email{boustani@uni-wuppertal.de}

\author{Alok Shukla}

\affiliation{Department of Physics, Indian Institute of Technology Bombay, Powai,
Mumbai 400076, India}

\email{shukla@phy.iitb.ac.in}

\providecommand{\tabularnewline}{\\}

\makeatother

\usepackage{babel}
\begin{document}
\begin{abstract}
In this work, we present large-scale electron-correlated computations
on various conformers of B$_{12}$H$_{12}$ and B$_{12}$H$_{6}$
clusters, to understand the reasons behind the high stability of di-anion
icosahedron ($I_{h}$) and cage-like B$_{12}$H$_{6}$ geometries.
Although the B$_{12}$-icosahedron is the basic building block in
some structures of bulk boron, it is unstable in its free form. Furthermore,
its H-passivated entity, i.e., B$_{12}$H$_{12}$ icosahedron is also
unstable in free form. However, dianion B$_{12}$H$_{12}$ has been
predicted to be stable as a perfect icosahedron in the free-standing
form. In order to capture the correct picture for the stability of
B$_{12}$H$_{12}^{-2}$ and B$_{12}$H$_{6}$ clusters, we optimized
these structures by employing the coupled-cluster singles-doubles
(CCSD) approach and cc-pVDZ basis set. We also performed vibrational
frequency analysis of the isomers of these clusters, using the same
level of theory to ensure the stability of the structures. For all
the stable geometries obtained from the vibrational frequency analysis,
we additionally computed their optical absorption spectra using the
time-dependent density functional theory (TDDFT) approach, at the
the B3LYP/6-31G{*} level of theory. Our calculated absorption spectra
could be probed in future experiments on these clusters.
\end{abstract}

\section{Introduction}

Boron, with one electron less than carbon, has unique bonding properties,
giving rise to several allotropes and compounds, with diverse physical
and electronic characteristics.\cite{boron-review} In contrast to
the solid carbon and carbon hydrides, which occur in nature abundantly
in the form of graphite, diamond, or hydrocarbons, solid boron as
well as boron hydrides (called boranes) do not exist in nature, and
must be synthesized in the lab. Besides $\gamma$-B$_{28}$ boron,
the well known boron solids $\alpha$- and $\beta$-rhombohedral boron
are composed of B$_{12}$ icosahedra. Even though the basic building
block unit in solid boron is a B$_{12}$ icosahedron, it is unstable
in isolated form due to the partly-filled, degenerate, highest-occupied
molecular orbitals (HOMOs). This leads to a symmetry reduction by
the Jahn-Teller-Effect, transforming the 3D icosahedral I$_{h}$-symmetry
into a quasi-planar 2D C$_{3v}$-symmetry, with six delocalized $\pi$-electrons,
analogous to benzene.\cite{pritam_jpcs_et_al} Nevertheless, the B$_{12}$
icosahedron plays a very important role in the boron chemistry, which
began to emerge in the early 20th century, with the efforts aimed
at exploring the possible synthesis of boranes, in analogy with the
hydrocarbons. By the middle of the 20th century, some boron hydride
compounds were synthesized exhibiting high thermal capacity, which
on burning could release huge quantities of heat, proving to be excellent
fuel sources, just like hydrocarbons. It is well known that the geometries
of boranes, i.e., boron hydrides of the form of B$_{x}$H$_{y}$,
are classified in closo, nido, and arachno structures. Furthermore,
Lipscomb proposed in 1992, that large closo boron hydrides forming
polyhedral boranes in analogy to carbon fullerenes, can be constructed.\cite{Lipscomb_et_al}
The most prominent and well known borane is the dodecahydro-closo-dodecaborate
B$_{12}$H$_{12}^{2-}$ di-anion, which also possesses an icosahedral
structure, but with an extremely high stability. This fact leads to
the question whether all structures of boranes depend on the content
of the hydrogen atoms and whether their chemical properties are tunable
through the number of the participating hydrogen atoms, and whether
they can affect and control the chemical reactivity. Ohishi \emph{et
al.}\cite{Ohishi_1_et_al,Ohishi_2_et_al,Ohishi_3_et_al} investigated
the formation of hydrogen-content-controlled B$_{12}$H$_{n}^{+}$
clusters experimentally through the decomposition and the ion-molecule
reaction of the decaborane B$_{10}$H$_{2m}^{+}$ (with $m=1-6$)
and diborane B$_{2}$H$_{6}$ molecules. They found that the detachment
of hydrogen atoms leads to an energy barrier for a structural transition.
They also confirmed that B$_{12}$H$_{n}^{+}$ clusters with more
hydrogen atoms ($n=7-12$) prefer the icosahedral structures, while
the clusters with fewer hydrogen atoms ($n=0-6$) favor planar structures.
Later on, Szwacki \emph{et al.}\cite{Szwacki_et_al} proposed a benzene-like
structure B$_{12}$H$_{6}$ called borozene, while Sahu and Shukla\cite{Sahu_et_al}
investigated the aromaticity of borozene, and computed its electronic
properties and optical absorption spectrum, along with the static
polarizability. Forte \emph{et al.}\cite{Forte_et_al} predicted larger
aromatic compounds composed of up to 26 borozene molecules in analogy
to the case of polycyclic aromatic hydrocarbons such as coronene and
coronene 19, for which the basic building block is benzene. To make
the analogy complete, they also reported that the dangling bonds of
these boron clusters are saturated by hydrogen atoms.\cite{Forte_et_al}

Here, we present a large-scale first-principles quantum-chemical study
of various isomers of B$_{12}$H$_{12}$ and B$_{2}$H$_{6}$ clusters,
with the motivation to investigate the reasons behind the high stability
of di-anion icosahedron ($I_{h}$) and cage-like B\textsubscript{12}H\textsubscript{6}
geometries. It is important to understand as to why the neutral B$_{12}$H$_{12}$-icosahedron
is unstable, while in some structures of bulk boron, B$_{12}$-icosahedron
is the basic unit. In order to understand the relationship between
the geometry and optical properties of these clusters, we also computed
the optical absorption spectra of all the stable isomers using the
TDDFT approach, at the  B3LYP/6-31G{*} level of theory. We believe
that ours is the first study of its kind probing the stability and
the optical properties of boranes, within a first-principles formalism.

\section{Theoretical approach and Computational details}

\label{sec:theory}

\begin{figure}
\begin{centering}
\includegraphics{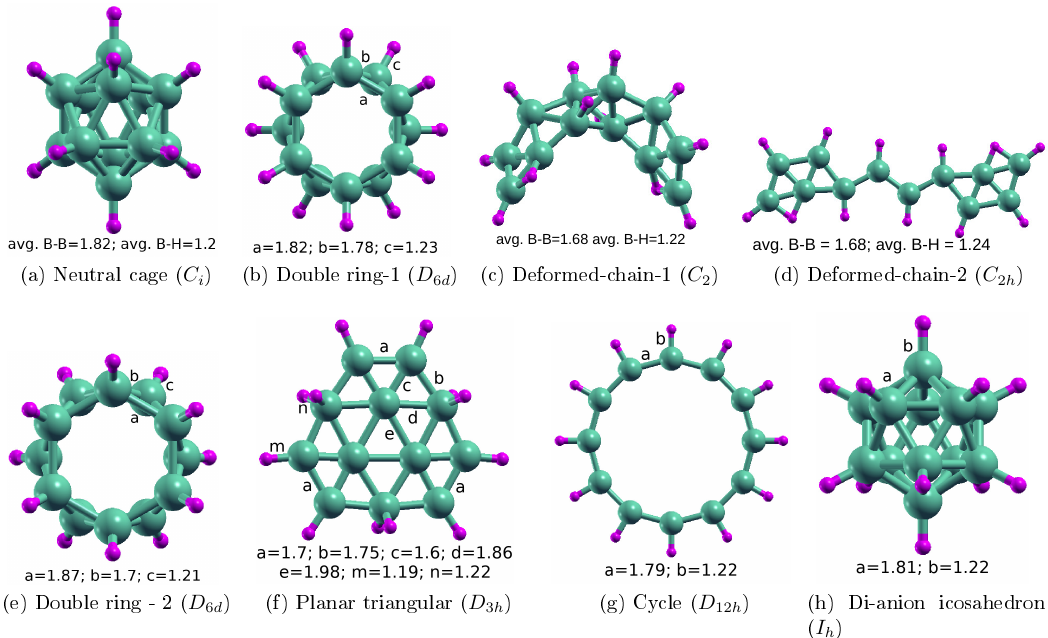}
\par\end{centering}
\caption{Optimized geometries of the various isomers of B$_{12}$H$_{12}$\protect\textsubscript{}
cluster, obtained using the CCSD approach, and the cc-pVDZ basis set.
The atoms with purple color indicate the hydrogen atoms, and the rest
of them are boron atoms. The point group symmetry of each isomer is
written in parentheses. All the bond lengths are in $\text{Å}$\ 
unit. \label{fig:optimized-geometries}}
\end{figure}

\begin{figure}
\begin{centering}
\includegraphics{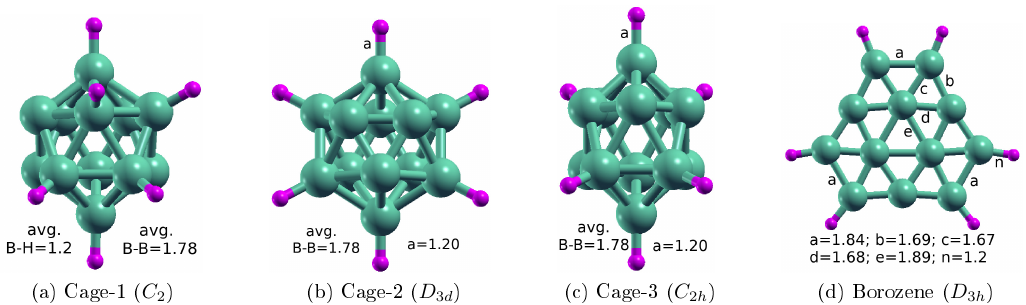}
\par\end{centering}
\caption{Optimized geometries of various isomers of B$_{12}$H$_{6}$\protect\textsubscript{}
cluster. Structures were optimized at the CCSD level of theory, employing
the cc-pVDZ basis sets. The atoms with purple color indicate the hydrogen
atoms, and the rest of them are boron atoms. The point group symmetry
of the isomers indicated inside parentheses. All the bond lengths
are in \AA\  unit. \label{fig:optimized-geometries-1}}
\end{figure}

We performed the geometry optimization of various isomers of B$_{12}$H$_{n}$
($n=6,\:12$) clusters by employing the coupled-cluster singles-doubles
(CCSD) approach and cc-pVDZ basis set. All the calculations were carried
out using the GAUSSIAN 16 computer program.\cite{g16} On examining
the many-electron wave functions of the final optimized states, we
found that the restricted closed-shell Hartree-Fock configuration
dominates the ground state wave functions of all the considered B$_{12}$H$_{12}$
and B$_{12}$H$_{6}$ isomers. Therefore, a single-reference electron-correlation
approach such as CCSD employing the HF state as reference, will yield
reasonable results. Single point energy was also computed at the obtained
optimized geometries, using the coupled-cluster singles-doubles with
perturbative triples (CCSD(T)) method, and the cc-pVTZ basis set to
improve the inclusion of electron-correlation effects. The optimized
geometries of various neutral isomers of B$_{12}$H$_{12}$ and B$_{12}$H$_{6}$
are presented in Figs. \ref{fig:optimized-geometries} and \ref{fig:optimized-geometries-1},
respectively. Additionally, Fig. \ref{fig:optimized-geometries} also
exhibits the optimized structure of dianion B$_{12}$H$_{12}^{-2}$.
The calculated ground state energy, correlation energy, and average
binding energy per atom, all computed using the CCSD(T) approach,
for each isomer are presented in the Table \ref{tab:total-energies}.
The correlation energy is the difference of energy computed at the
CCSD(T) and the Hartree-Fock energy for a given system. The binding
energy which can ensure the stability of a system was calculated using
the following formula 

\begin{equation}
\frac{E_{BE}}{N}=\frac{N_{B}}{N}E_{B}+\frac{N_{H}}{N}E_{H}-\frac{E_{T}}{N},\label{eq:BE}
\end{equation}

where $N_{B}$ and $N_{H}$ are the respective numbers of boron and
hydrogen atoms present in the system, while $E_{B}$ and $E_{H}$
are the energies of single boron and hydrogen atoms, respectively.
$N$ is the total number of atoms present in the system, whereas $E_{T}$
is the total energy of the system. Here, $E_{B}$ and $E_{H}$ were
taken to be -24.598101 Hartree, and -0.4998098 Hartree, respectively,
computed using the CCSD(T) method, and the cc-pVTZ basis set. 

Furthermore, we performed the vibrational frequency analysis of the
conformers of B\textsubscript{12}H\textsubscript{12} and B\textsubscript{12}H\textsubscript{6}
clusters using the CCSD approach, along with the cc-pVDZ basis set
to verify their stability. We find that all the isomers of B\textsubscript{12}H\textsubscript{12}
cluster are unstable, while those of the B\textsubscript{12}H\textsubscript{6}
cluster are stable. To further confirm our results on the instability
of B\textsubscript{12}H\textsubscript{12} isomers , we also optimized
and calculated their vibrational frequencies using the density-functional
theory (DFT), and all the isomers, except a \textquotedbl deformed
chain\textquotedbl{} structure, were found to be unstable.

Next, we computed the optical absorption spectra of the stable structures
of all the clusters using the time-dependent density functional theory
(TDDFT) approach. For the purpose, we employed the hybrid exchange-correlation
functional, B3LYP,\cite{Becke_TDDFT,Lee_Yang_Parr,Stephens_TDDFT}
and a localized Gaussian basis set, 6-31+G{*}, which is composed of
a valance double-$\zeta$ set, augmented with s and p diffuse functions,
and d polarization functions for the non-hydrogen atoms. Although,
the B3LYP hybrid functional was mainly developed for ground-state
calculations, but it has also been found to yield accurate results
on the excited states of both closed-shell,\cite{Bauernschmitt_et_al}
and open-shell systems.\cite{Hirata_et_al,Hirata_2_et_al} In some
of the recent studies, it has been found that the above computational
framework (B3LYP\textbackslash 6-31+G{*}) for both, ground and excited
states led to high-accuracy results, in excellent agreement with the
experiments for the $\pi$-conjugated aromatic hydrocarbon molecules.\cite{pritam_jpcc,Malloci_et_at,Malloci_2_et_at,Mocci_et_al}
We adopted the TDDFT approach to compute the optical properties of
the stable structures, because of the two primary reasons: (i) earlier
studies exhibit reliable performance, and (ii) computing the optical
absorption using wave-function-based electron-correlated approaches
is computationally quite prohibitive, for the relatively larger clusters
considered in this work. Therefore, we performed the ground state
calculations using the most accurate electron-correlated approach
(coupled-cluster approach), while the excited-state calculations were
carried out by employing the TDDFT methodology.

\begin{table}
\caption{Total energies, and related data for various isomers of B\protect\textsubscript{12}H\protect\textsubscript{12}
cluster including its di-anion, and B\protect\textsubscript{12}H\protect\textsubscript{6}
cluster, are presented below. The point group symmetry of each isomer
is shown next to its name. The ground state energies (in Hartree),
computed at the CCSD(T) level, and the relative energies (in eV) with
respect to their lowest-energy isomers, are also presented. The difference
in the total energies of a system at the CCSD(T) and the Hartree-Fock
levels of theory, employing cc-pVTZ basis functions, is defined as
the correlation energy of that system. The average (avg.) binding
energies per atom (in eV) are listed in the last column to indicate
the stability of various structures. \label{tab:total-energies}}

\begin{tabular}{ccccccc}
\hline 
 &  & Point  & Ground state & Relative & Correlation  & Avg.Binding \tabularnewline
Cluster & Isomer & group & energy (Ha) & energy (eV) & energy (eV) & energy per\tabularnewline
 &  &  &  &  &  & atom (eV)\tabularnewline
\hline 
\hline 
 &  &  &  &  &  & \tabularnewline
B\textsubscript{12}H\textsubscript{12} & Neutral Cage & C\textsubscript{i} & -304.6848659 & 0.00 & 43.44 & 3.98\tabularnewline
 &  &  &  &  &  & \tabularnewline
 & Double ring - 1 & D\textsubscript{6d} & -304.5693208 & 3.14 & 43.2 & 3.85\tabularnewline
 &  &  &  &  &  & \tabularnewline
 & Deformed-chain-1 & C\textsubscript{2} & -304.5368726 & 4.03 & 41.17 & 3.81\tabularnewline
 &  &  &  &  &  & \tabularnewline
 & Double ring - 2 & D\textsubscript{6d} & -304.5010969 & 5.00 & 46.56 & 3.77\tabularnewline
 &  &  &  &  &  & \tabularnewline
 & Planar triangular & D\textsubscript{3h} & -304.4501781 & 6.39 & 42.48 & 3.71\tabularnewline
 &  &  &  &  &  & \tabularnewline
 & Deformed-chain-2 & C\textsubscript{2h} & -304.3491315 & 9.14 & 40.55 & 3.60\tabularnewline
 &  &  &  &  &  & \tabularnewline
 & Cycle & D\textsubscript{12h} & -303.9550167 & 19.86 & 35.04 & 3.15\tabularnewline
 &  &  &  &  &  & \tabularnewline
B\textsubscript{12}H\textsubscript{12}\textsuperscript{-2} & Di-anion Icosahedron & I\textsubscript{h} & -304.9486026 & 0.00 & 42.97 & 4.28\tabularnewline
 &  &  &  &  &  & \tabularnewline
B\textsubscript{12}H\textsubscript{6} & Cage-1 & C\textsubscript{2} & -300.9978675 & 0.00 & 39.42 & 4.27\tabularnewline
 &  &  &  &  &  & \tabularnewline
 & Cage-2 & D\textsubscript{3d} & -300.9903705 & 0.20 & 39.44 & 4.25\tabularnewline
 &  &  &  &  &  & \tabularnewline
 & Cage-3 & C\textsubscript{2h} & -300.9490125 & 1.33 & 39.08 & 4.19\tabularnewline
 &  &  &  &  &  & \tabularnewline
 & Borozene & D\textsubscript{3h} & -300.9352434 & 1.70 & 37.26 & 4.17\tabularnewline
 &  &  &  &  &  & \tabularnewline
\hline 
\end{tabular}
\end{table}

\section{Results and Discussion}

\label{sec:results}

In this section, we present and discuss the results of our calculations
for each conformer of B$_{12}$H$_{n}$ ($n=6,12$) clusters considered
in this work, and also the optical properties of the low-lying stable
isomers of these clusters. For the $n=12$ case, in addition to the
neutral clusters, we also discuss the case of its di-anion.

\subsection{B\protect\textsubscript{12}H\protect\textsubscript{12} Isomers}

\subsubsection{Cage \label{subsec:Neutral-Icosahedron}}

\begin{figure}
\begin{centering}
\includegraphics[scale=0.7]{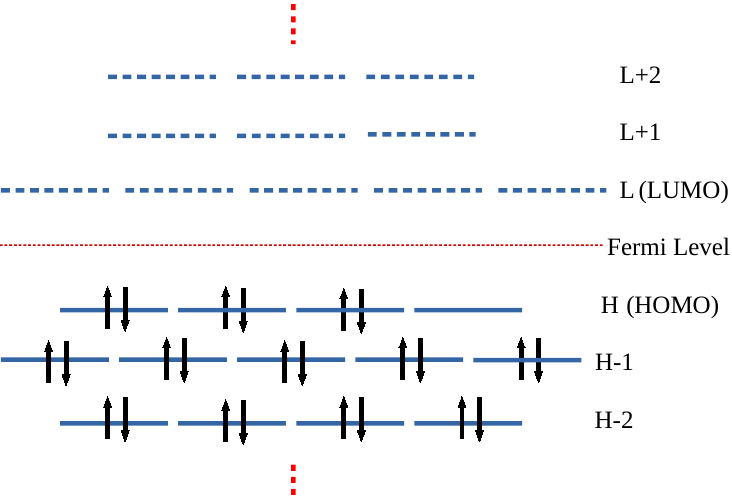}
\par\end{centering}
\caption{Energy levels of the B$_{12}$H$_{12}$ cage of perfect icosahedral
symmetry ($I_{h}$ point group), computed using the STO-3G basis,
and no electrons. Because the HOMO is four-fold degenerate, there
are not enough electrons in the molecule to form a closed-shell structure,
leading to its instability.\label{fig:b12h12-energy-levels}}
\end{figure}

In none of our calculations performed on the cage structures was the
optimized geometry found to be a perfect icosahedron, with the $I_{h}$
point group. However, the geometry optimization performed using the
CCSD approach using the cc-pVDZ basis set, predicted a low-symmetry
cage with the $C_{i}$ point group as shown in Fig. \ref{fig:optimized-geometries}(a)
to have the lowest energy of all the isomers considered. In this structure,
the average B-H and B-B bond lengths are 1.2 \AA~ and 1.82 \AA~,
respectively. But, when we performed the vibrational frequency analysis
also at the CCSD/cc-pVDZ level of theory, we obtained five imaginary
frequencies implying that this structure is completely unstable. We
note that long ago Longuet-Higgins and Roberts\cite{Longuet-Higgins},
based on a simple tight-binding type molecular orbital approach, had
also predicted the cage-B$_{12}$H$_{12}$ with the $I_{h}$ symmetry
to be unstable. Their argument was based on the fact that due to high-symmetry
of the structure, the $I_{h}$-cage will have degenerate HOMOs leading
to an open-shell structure, and hence the structural instability.
In Fig. \ref{fig:b12h12-energy-levels} we present the energy levels
of cage-B$_{12}$H$_{12}$ with the perfect $I_{h}$ symmetry, along
their electron filling according to the aufbau principle. From the
figure it is obvious that due to the four-fold degeneracy, the molecule
will have a partially filled HOMO, making it a candidate for Jahn-Teller
distortion. However, our geometry optimization calculations go beyond
that and have revealed that cage structure of B$_{12}$H$_{12}$,
even with a lower symmetry ($C_{i}$), is unstable.

\subsubsection{Deformed chain\textcolor{red}{{} \label{subsec:Deformed-chain}}}

The neutral chain-type isomer of B$_{12}$H$_{12}$ considered in
this work is named as ``deformed chain'', because of its three-dimensional
structure similar to that of a curved (and not straight) chain (see
Fig.\textcolor{red}{{} }\ref{fig:optimized-geometries}(c)) The geometry
of this isomer called deformed-chain-1 shown in Fig.\textcolor{red}{{}
}\ref{fig:optimized-geometries}(c) was optimized using the DFT, in
which we employed the B3LYP exchange-correlation functional, along
with the 6-31+G{*} basis set, and the final structure belongs to the
$C_{2}$ point group. We also performed the vibrational frequency
analysis at the same level of theory, which predicted this structure
to be a stable one. The average nearest-neighbor B-B distance is 1.68
$\text{Å}$, while the B-H bond length is 1.22 $\text{Å}$.

However, when we performed the geometry optimization at the CCSD/cc-pVDZ
level of theory, we obtained a totally different deformed-chain structure
called deformed-chain-2, as depicted in Fig. \ref{fig:optimized-geometries}(d),
with a planar geometry, and $C_{2h}$ point group. This isomer has
average B-B and B-H bond lengths of 1.68 $\text{Å}$, and 1.24 $\text{Å}$
, respectively. However, on performing vibrational frequency analysis
at the same level of theory, we found that the structure is unstable. 

\begin{figure}
\begin{centering}
\textcolor{red}{\includegraphics[scale=0.4]{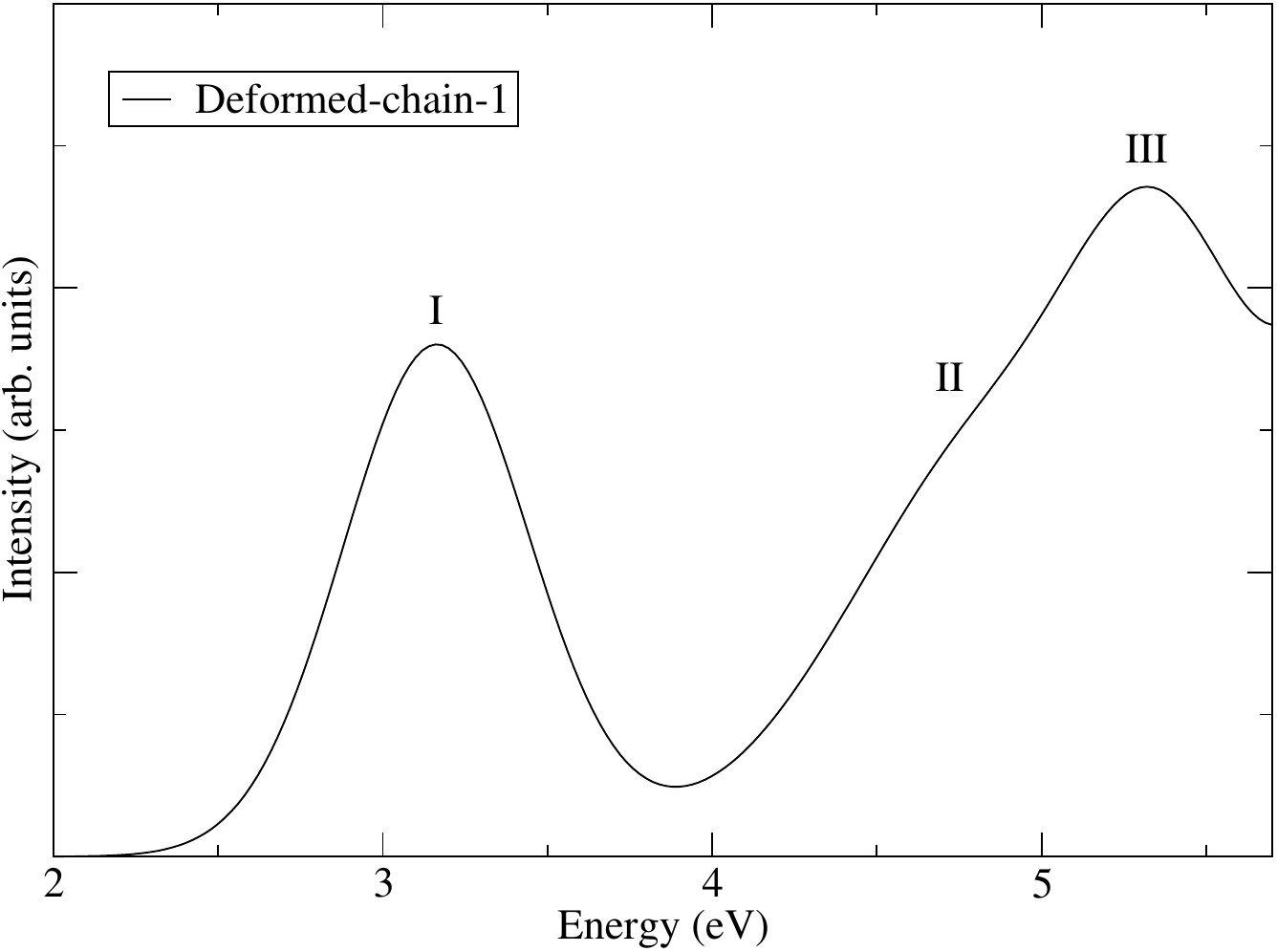}}
\par\end{centering}
\textcolor{red}{\caption{Optical absorption spectrum of the deformed-chain-1 (B$_{12}$H$_{12}$)
isomer, computed using the TDDFT approach, at the B3LYP/6-31+G{*}
level of theory.\label{fig:optics-deformed-chain}}
}
\end{figure}

Because deformed-chain-1 (Fig.\textcolor{red}{{} }\ref{fig:optimized-geometries}(c))
was found to be stable at the DFT-B3LYP level of theory, we also computed
its optical absorption spectrum using the TDDFT approach, at the same
level of theory, and the results are presented in Fig.\textcolor{red}{{}
}\ref{fig:optics-deformed-chain} and Table \ref{tab:wfs-deformed-chain}. 

\textcolor{red}{}
\textcolor{red}{}
\begin{table}
\textcolor{red}{\caption{Excitation energies of the peaks with significant intensities in the
optical absorption spectrum of the deformed-chain-1 isomer of B$_{12}$H$_{12}$
presented in Fig. \ref{fig:optics-deformed-chain}, along with the
corresponding oscillator strengths (OS). Additionally, the main configurations
contributing to the many-body wave functions of the excited states
corresponding to the peaks of the absorption spectrum.\label{tab:wfs-deformed-chain}}
}
\centering{}%
\begin{tabular}{|c|c|c|c|}
\hline 
Peak & Energy (eV) & OS & Wave function\tabularnewline
\hline 
\hline 
I & 3.16 & 0.222 & $\arrowvert H\rightarrow L\rangle$\tabularnewline
\hline 
 &  &  & \tabularnewline
\hline 
II & 4.81 & 0.121 & $\arrowvert H\rightarrow L+2\rangle$\tabularnewline
\hline 
 &  &  & \tabularnewline
\hline 
III & 5.36 & 0.097 & $\arrowvert H\rightarrow L+5\rangle$\tabularnewline
\hline 
\end{tabular}
\end{table}

From the excited-state wave functions presented in Table \ref{tab:wfs-deformed-chain},
it is obvious that the peaks appearing in the absorption spectrum
are mainly due to the single excitations from the HOMO of the isomer.
The absorption starts with a intense peak at 3.16 eV, dominated by
the transition $\arrowvert H\rightarrow L\rangle$. It is followed
by a shoulder peak near 4.80 eV due to the transition $\arrowvert H\rightarrow L+2\rangle$,
while the most intense peak near 5.36 eV is characterized by the transition
$\arrowvert H\rightarrow L+5\rangle$. All the orbitals of this isomer
are non-degenerate.

\subsubsection{Double ring - 1 }

This conformer consists of two hexagonal rings of boron (B) atoms
kept on top of each other, but in a staggered configuration with the
relative twist angle of almost 30$^{\circ}$, and each boron atom
bonded with one hydrogen atom as shown in Fig. \ref{fig:optimized-geometries}(b).
This isomer has D\textsubscript{6d} point group symmetry, with its
total energy 3.14 eV higher as compared to the neutral cage conformer.
The distances between the two closest in-plain and out-of-plain B-atoms
are 1.82 and 1.78 $\text{Å}$, whereas the uniform B-H bond length
of the geometry is 1.23 $\text{Å}$.\textcolor{blue}{{} }The perpendicular
distance between the planes containing two rings is 1.51 $\text{Å}$.
The H-B-B-H dihedral angle, where the B- and H-atoms are closest as
well as out-of-plain atoms, is 1.08$^{\circ}$. Though the B-B bond
length of this isomer is larger as compared to our earlier studied
stable structure of the bare B\textsubscript{12}-double ring, the
point group symmetry ($D_{6d}$) remains unchanged.\cite{pritam_jpcs_et_al}
This geometry was found to be unstable both at the CCSD/ cc-pVDZ as
well as B3LYP/6-31+G{*} levels of theory.

\subsubsection{Double ring - 2 }

This isomer has a very similar structure as double-ring - 1 isomer
discussed above, with a larger in-plain B-B atom bond length of 1.87
$\text{Å}$ and shorter out-of-plain B-B atom distance of 1.70 $\text{Å}$
presented in Fig. \ref{fig:optimized-geometries}(e). The uniform
B-H bond length of this isomer has also reduced, in comparison, to
1.21 $\text{Å}$. While both the double ring isomers of the B$_{12}$H$_{12}$
cluster have D\textsubscript{6d} point group symmetry, but the H-B-B-H
dihedral angle of this isomer is comparatively larger (59.46$^{\circ}$).
This isomer is predicted to be 5.00 eV higher in energy as compared
to the neutral cage structure. The perpendicular distance between
the planes containing the two rings is 1.40 $\text{Å}$.\textcolor{red}{{}
}This structure was also found to be unstable because the vibrational
analysis performed at the CCSD as well as DFT (B3LYP) levels of theory
led to imaginary frequencies. 

\subsubsection{Planar triangular }

This conformer is about 6.39 eV higher in energy as compared to the
neutral cage isomer, with a D$_{3h}$ point group symmetry. It is
composed of an inner-ring of three boron atoms, an outer-ring of nine
boron atoms, six in-plain hydrogen atoms, and six out-of-plain hydrogen
atoms. From Fig. \ref{fig:optimized-geometries}(f), it is evident
that the outer-ring consists of two different bond lengths, \emph{i.e.},
1.70 $\text{Å}$ and 1.75 $\text{Å}$, while the inner-ring is composed
of one type of bond length (1.98 $\text{Å}$). The connecting bonds
between the outer-ring and the inner-ring boron atoms are of two types,
\emph{i.e.}, 1.60 $\text{Å}$ and 1.86 $\text{Å}$. For the six in-plain
hydrogen atoms, the B-H bond length is 1.19 $\text{Å}$, but for the
six out-of-plain hydrogen atoms, the B-H bond distance is found to
be 1.22 $\text{Å}$. The bare B$_{12}$ - quasi-planar isomer which
we studied earlier\cite{pritam_jpcs_et_al}, also has a similar structure,
except that three boron atoms of the inner-ring are out of the plain,
whereas all the boron atoms of this structure are in the same plain.
However, both CCSD as well as DFT-B3LYP calculations predict this
isomer to be unstable.

\subsubsection{Cycle}

This isomer has a completely planar geometry with the $D_{12h}$ point-group
symmetry, and the structure is presented in Fig. \ref{fig:optimized-geometries}(g).
It is 19.86 eV higher in energy as compared to the neutral cage structure.
Clearly, the twelve boron atoms are placed on a circle, with each
hydrogen atom attached to a boron atom in the same plain. The uniform
bond length between two consecutive boron atoms is found to be 1.79
$\text{Å}$, while the bond distance between the boron and hydrogen
atoms is 1.22 $\text{Å}$. But, the vibrational analysis based calculations
performed both using the CCSD and the DFT-B3LYP approaches predicted
this isomer also to be unstable.

\subsection{Di-anion Icosahedron (B\protect\textsubscript{12}H\protect\textsubscript{12}\protect\textsuperscript{-2})\label{subsec:Di-anion-cage-BH12}}

In Fig. \ref{fig:b12h12-energy-levels} we saw that the neutral B$_{12}$H$_{12}$
was short of a closed-shell structure by two electrons. Therefore,
the question arises whether perfect icosahedral dianion B$_{12}$H$_{12}^{-2}$
will be stable? Longuet-Higgins and Roberts\cite{Longuet-Higgins}
had indeed speculated this ion to be stable, because of its closed-shell
structure. Later on, Pitochelli and Hawthorne experimentally isolated
the dianion B$_{12}$H$_{12}^{-2}$, and identified it as having the
icosahedral structure.\cite{Pitochelli_et_al} 

To probe this matter, we performed CCSD-based geometry optimization
on the cage structure of B$_{12}$H$_{12}^{-2}$, and obtained the
final geometry shown in Fig. \ref{fig:optimized-geometries}(h), which
has a perfect icosahedral shape with the $I_{h}$ point group symmetry.
As shown in the figure, the optimized B-B bond length is 1.81 \AA,
whereas the B-H bond lenght is 1.22 \AA. We also performed vibrational
frequency analysis on this structure using again CCSD/cc-pVDZ level
of theory, and obtained all the frequencies to be real, confirming
the stability of this perfect icosahedral structure of B$_{12}$H$_{12}^{-2}$.
Additionally, we also computed the energies of other geometries of
B$_{12}$H$_{12}^{-2}$, and found the energy of the $I_{h}$ structure
to be the lowest. We also computed the first ionization potential
of the icosahedral B$_{12}$H$_{12}^{-2}$\textcolor{red}{{} }at the
CCSD(T)/cc-pVTZ level of theory, and the calculated value of 2.03
eV suggests that the molecule is fairly stable against the ionization
as well.

\begin{figure}
\begin{centering}
\includegraphics[scale=0.4]{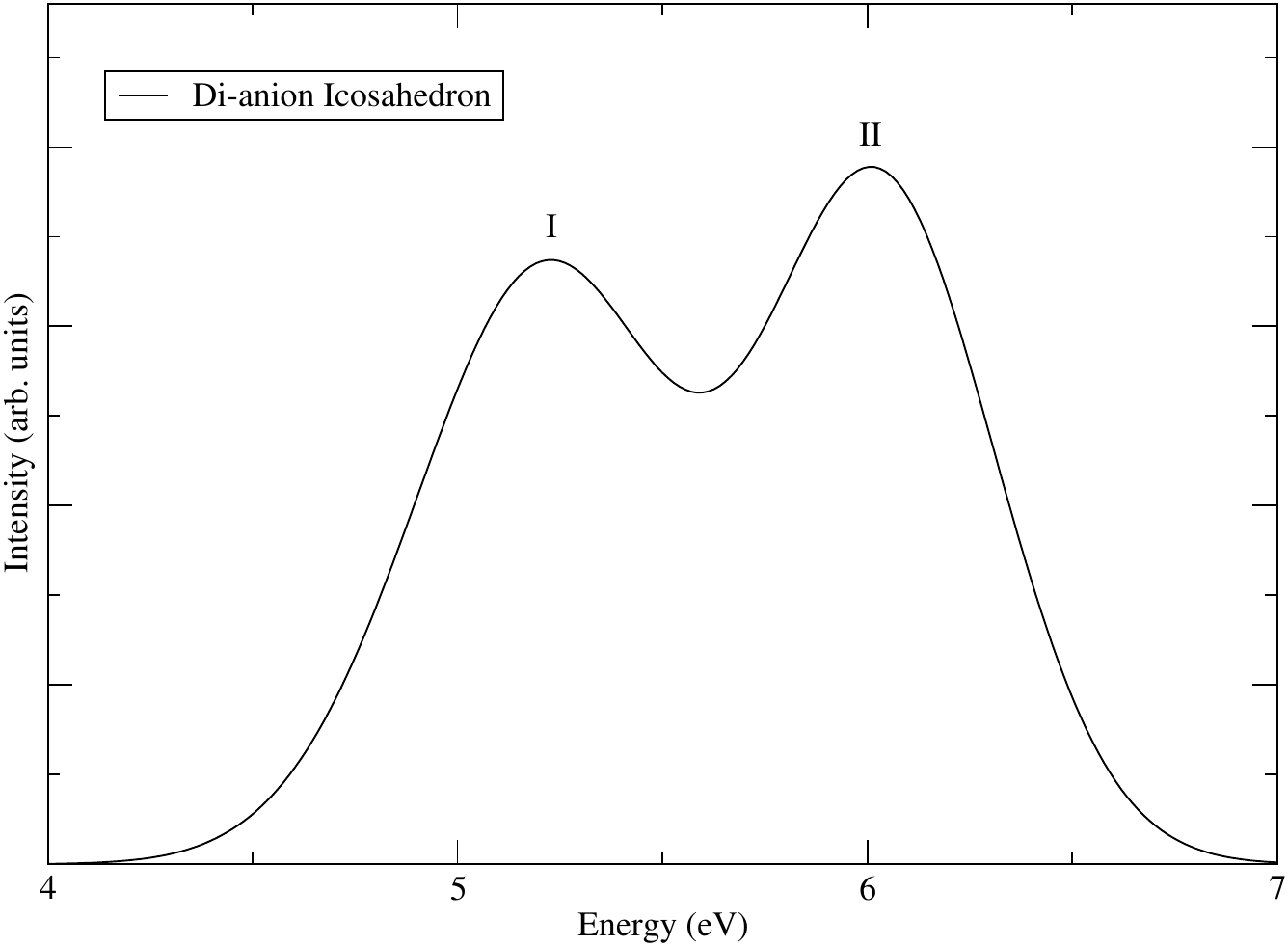}
\par\end{centering}
\caption{The optical absorption spectrum of di-anion icosahedron isomer calculated
using the TDDFT approach at B3LYP/6-31+G{*} level of theory.\label{fig:optics-bh12-ih}}
\end{figure}
Given the stability of this molecule, we computed its photo-absorption
spectrum at the B3LYP/6-31+G{*} level of theory, the results of which
are presented in Fig. \ref{fig:optics-bh12-ih}. The descriptions
of the excitations contributing to various peaks are presented in
Table \ref{tab:wfs-bh12-ih}. From the figure and the table it is
obvious that the absorption spectrum of this conformer mainly consist
of two intense peaks located near 5 eV and 6 eV. Given the fact that
both the peaks are at significantly higher energies than the first
ionization potential of the ion (2.03 eV), we conclude that it will
be impossible to measure these peaks experimentally.

\begin{table}
\caption{The locations of the peaks with significant intensities in the optical
absorption spectrum of the di-anion icosahedral B$_{12}$H$_{12}^{-2}$
computed using the TDDFT approach, along with the corresponding oscillator
strengths (OS), and the wave functions of the excited states. \label{tab:wfs-bh12-ih}}

\centering{}%
\begin{tabular}{|c|c|c|c|}
\hline 
Peak & Energy (eV) & OS & Wave function\tabularnewline
\hline 
\hline 
I & 4.97 & 0.034 & $\arrowvert H\rightarrow L+2\rangle$\tabularnewline
\hline 
 & 5.26 & 0.114 & $\arrowvert H-1\rightarrow L+1\rangle$\tabularnewline
\hline 
II & 6.03 & 0.157 & $\arrowvert H-1\rightarrow L+3\rangle$\tabularnewline
\hline 
\end{tabular}
\end{table}

\subsection{B\protect\textsubscript{12}H\protect\textsubscript{6} Isomers}

In this section we discuss the ground-state geometries of four stable
isomers of B$_{12}$H$_{6}$, optimized at the CCSD level of theory.
We note that out of the four, three conformers have cage-like structures,
while the fourth one known as borozene, has a perfectly planar geometry.
Furthermore, we found all the four conformers to be vibrationally
stable. Therefore, to study their optical properties, we also computed
their photoabsorption spectra using the TDDFT approach.

\subsubsection{Cage - 1 }

This lowest energy conformer of the B$_{12}$H$_{6}$ cluster belongs
to the $C_{2}$ point group, and has a cage-type appearance presented
in Fig \ref{fig:optimized-geometries-1}(a). Six of the twelve boron
atoms are bonded with one hydrogen atom each, with the average B-H
bond length 1.20$\text{Å}$. The average B-B bond distance is found
to be 1.78 $\text{Å}$. As discussed in our earlier work,\cite{pritam_jpcs_et_al}
the HOMO orbital of the isolated icosahedral B\textsubscript{12}
structure has a five-fold degeneracy, of which three orbitals are
unoccupied, thus making it unstable against the Jahn-Teller distortion.
However, the if we add six more electrons to the system by attaching
six hydrogen atoms, the resultant B$_{12}$H$_{6}$ cage, will be
stable against the distortion because of its closed-shell structure.
Indeed, the vibrational frequency analysis performed at the CCSD/cc-pVDZ
level of theory predicts that this isomer has a stable structure,
without any imaginary frequency.

\begin{figure}
\begin{centering}
\includegraphics[scale=0.4]{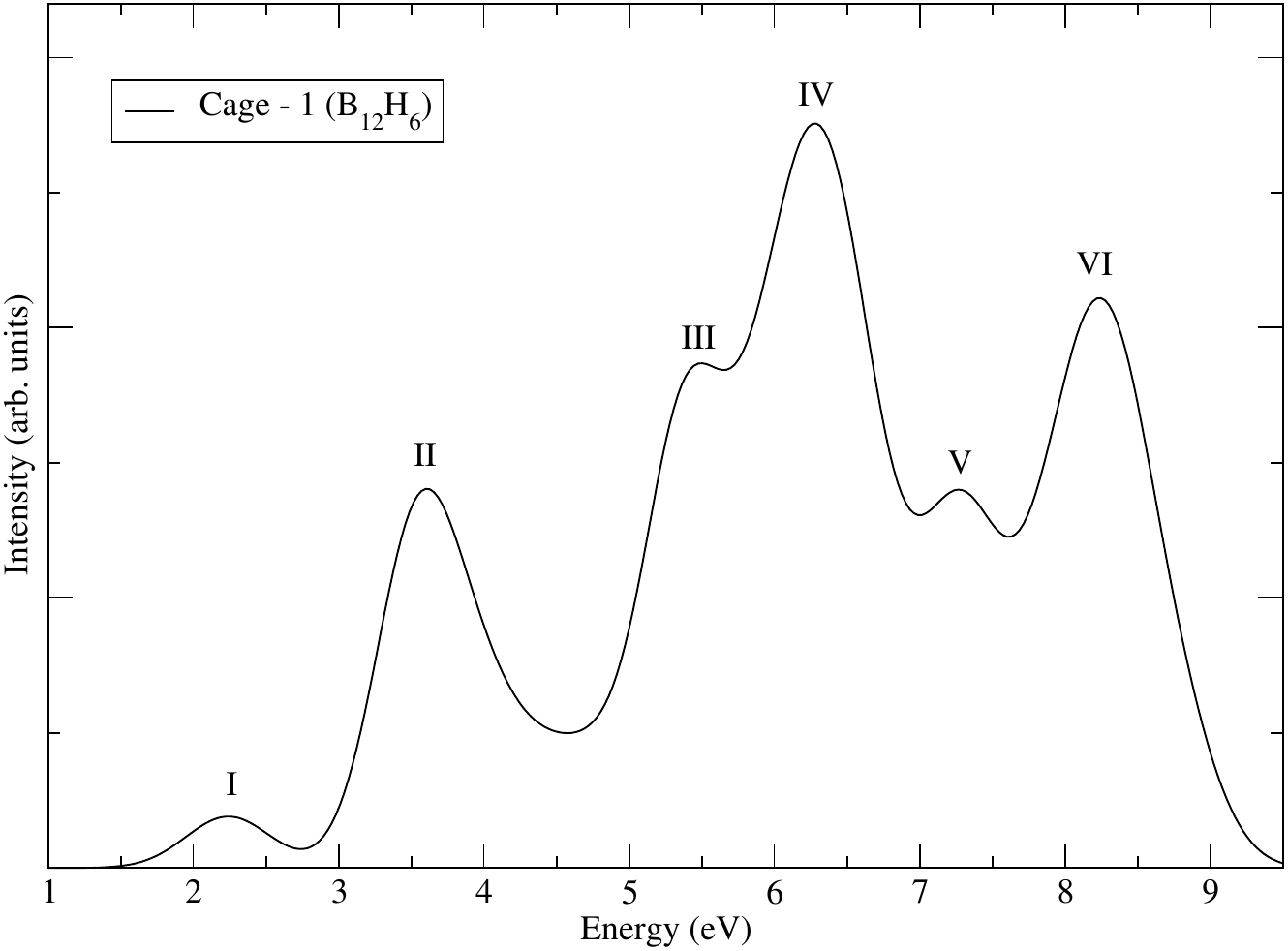}
\par\end{centering}
\caption{The optical absorption spectrum of cage-1 (B$_{12}$H$_{6}$) isomer
calculated using the TDDFT approach at B3LYP/6-31+G{*} level of theory.\label{fig:optics-cage-b12h6}}
\end{figure}

\begin{table}
\caption{Peak locations in the optical absorption spectrum of cage-1 isomer
of the B$_{12}$H$_{6}$ cluster, computed using the TDDFT approach,
along with the corresponding oscillator strength (OS), and the wave
functions of the excited states contributing to the peaks. \label{tab:wfs-cage-b12h6}}

\centering{}%
\begin{tabular}{|c|c|c|c|}
\hline 
Peak & Energy (eV) & OS & Wave function\tabularnewline
\hline 
\hline 
I & 2.18 & 0.013 & $\arrowvert H\rightarrow L+2\rangle$\tabularnewline
\hline 
 & 2.31 & 0.011 & $\arrowvert H\rightarrow L+1\rangle$\tabularnewline
\hline 
II & 3.52 & 0.135 & $\arrowvert H\rightarrow L+3\rangle$\tabularnewline
\hline 
III & 5.35 & 0.120 & $\arrowvert H\rightarrow L+4\rangle$\tabularnewline
\hline 
IV & 6.16 & 0.067 & $\arrowvert H-9\rightarrow L+1\rangle$\tabularnewline
\hline 
 & 6.52 & 0.065 & $\arrowvert H\rightarrow L+16\rangle$\tabularnewline
\hline 
V & 7.28 & 0.064 & $\arrowvert H\rightarrow L+19\rangle$\tabularnewline
\hline 
VI & 8.29 & 0.053 & $\arrowvert H-1\rightarrow L+12\rangle$\tabularnewline
\hline 
\end{tabular}
\end{table}
 The calculated optical absorption spectrum of this isomer is shown
in Fig. \ref{fig:optics-cage-b12h6}, whereas information related
to various peaks is represented in Table \ref{tab:wfs-cage-b12h6}.
It is obvious from the table that several peaks occur due to transition
from the HOMO to the virtual orbitals. The absorption spectrum starts
with a small peak near 2.20 eV due to two close-lying excited states
whose wave functions consist mainly of the single excitation $\arrowvert H\rightarrow L+2\rangle$
and $\arrowvert H\rightarrow L+1\rangle$. It is followed by two moderately
intense peaks at 3.52 and 5.35 eV that are dominated by the transitions
$\arrowvert H\rightarrow L+3\rangle$ and $\arrowvert H\rightarrow L+4\rangle$,
respectively. The most intense peak near 6.50 eV is due to two excited
states characterized by the transitions $\arrowvert H-9\rightarrow L+1\rangle$
and $\arrowvert H\rightarrow L+16\rangle$. The peaks at 7.28 and
8.29 eV are largely composed of the singly excitations $\arrowvert H\rightarrow L+19\rangle$
and $\arrowvert H-1\rightarrow L+12\rangle$, respectively. All the
molecular orbitals involved in optical transitions of this isomer
are non-degenerate. 

\subsubsection{Cage - 2 }

\begin{figure}
\begin{centering}
\includegraphics[scale=0.4]{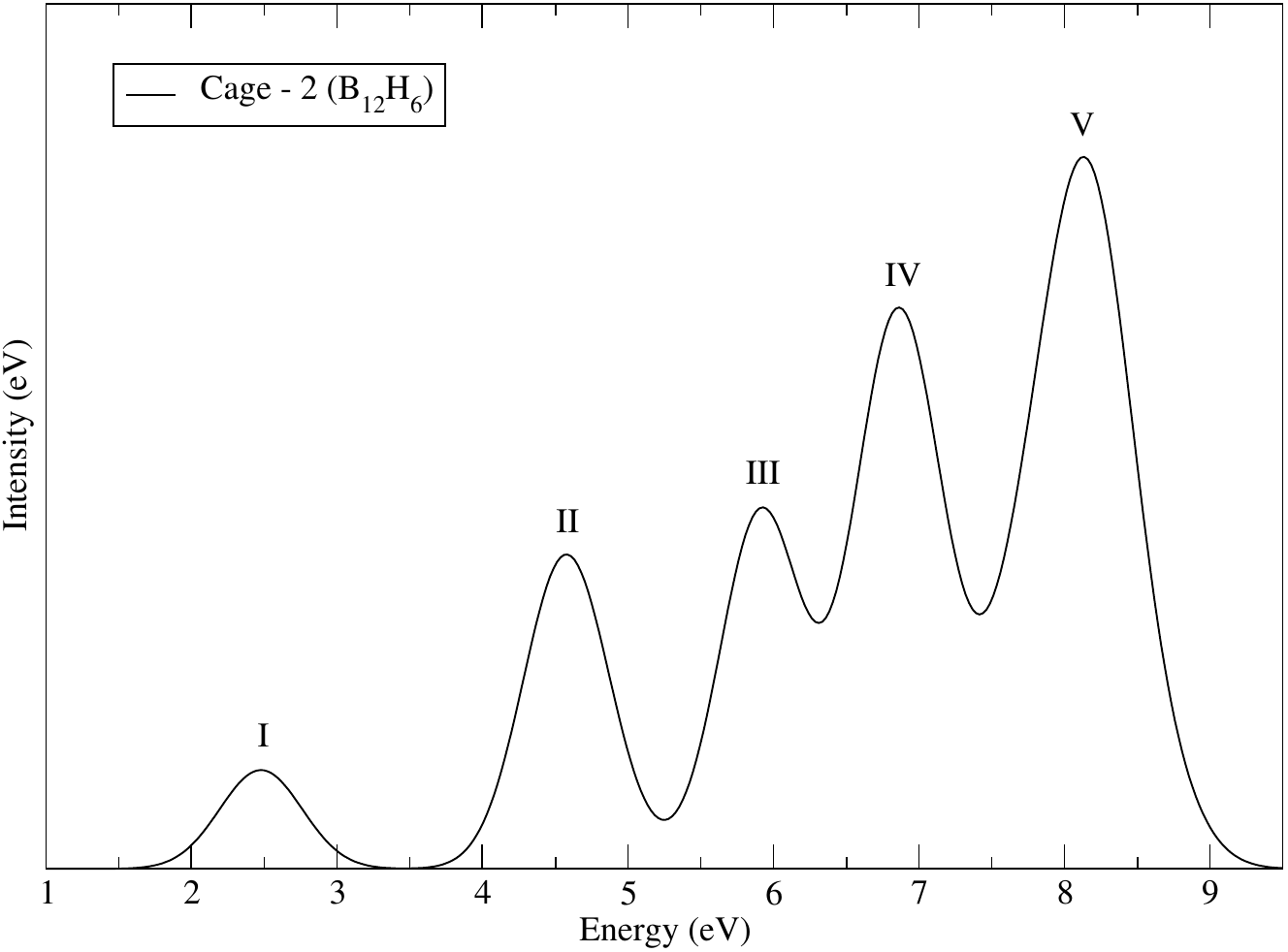}
\par\end{centering}
\caption{The optical absorption spectrum of cage-2 B$_{12}$H$_{6}$ isomer
calculated using the TDDFT approach at B3LYP/6-31+G{*} level of theory.\label{fig:optics-cage-2-b12h6}}
\end{figure}

This isomer of the B$_{12}$H$_{6}$ cluster presented in Fig. \ref{fig:optimized-geometries-1}(b),
which also has a cage geometry with a $D_{3d}$ point group symmetry,
is only 0.2 eV higher than its lowest-energy conformer. The average
optimized lengths of the B-B and B-H bonds are 1.78 $\text{Å}$, and1.20
$\text{Å}$, respectively. According to the CCSD/cc-pVDZ level vibrational
frequency analysis, this isomer is completely stable. Therefore, we
calculated its optical absorption spectrum shown in Fig. \ref{fig:optics-cage-2-b12h6},
while the dominant configurations contributing to its peaks are presented
in Table \ref{tab:wfs-cage-2-b12h6}.

\begin{table}
\caption{Peak locations of the optical absorption spectrum of cage-2 isomer
of the B$_{12}$H$_{6}$ computed using the TDDFT approach, along
with the corresponding oscillator strength (OS), and the wave functions
of the excited states contributing to its peaks.\label{tab:wfs-cage-2-b12h6}}

\centering{}%
\begin{tabular}{|c|c|c|c|}
\hline 
Peak & Energy (eV) & OS & Wave function\tabularnewline
\hline 
\hline 
I & 2.48 & 0.028 & $\arrowvert H-1\rightarrow L\rangle$\tabularnewline
\hline 
II & 4.53 & 0.070 & $\arrowvert H\rightarrow L+1\rangle$\tabularnewline
\hline 
III & 5.92 & 0.201 & $\arrowvert H\rightarrow L+3\rangle$\tabularnewline
\hline 
IV & 6.75 & 0.130 & $\arrowvert H-1\rightarrow L+6\rangle$\tabularnewline
\hline 
V & 7.80 & 0.096 & $\arrowvert H-9\rightarrow L\rangle$\tabularnewline
\hline 
 & 8.19 & 0.120 & $\arrowvert H-1\rightarrow L+12\rangle$\tabularnewline
\hline 
\end{tabular}
\end{table}
 The absorption begins with a small peak at 2.48 eV due to the transition
$\arrowvert H-1\rightarrow L\rangle$, where $\arrowvert H-1\rangle$
is non-degenerate but $\arrowvert L\rangle$ is doubly degenerate.
Peak II appears due to the state whose wave function is largely composed
of the $\arrowvert H\rightarrow L+1\rangle$ transition. The $\arrowvert L+1\rangle$
is also doubly degenerate. It is followed by two moderately intense
peaks at 5.92 and 6.75 eV characterized by the transitions $\arrowvert H\rightarrow L+3\rangle$
and $\arrowvert H-1\rightarrow L+6\rangle$, respectively. The most
intense peak of the spectrum near 8 eV is due to two excited states,
dominated by the single excitations $\arrowvert H-9\rightarrow L\rangle$
and $\arrowvert H-1\rightarrow L+12\rangle$, where both $\arrowvert H-9\rangle$
and $\arrowvert L+12\rangle$ are doubly degenerate. This most intense
peak is also associated with the orbitals, which are far away from
the Fermi level.

\subsubsection{Cage - 3 }

\begin{figure}
\begin{centering}
\includegraphics[scale=0.4]{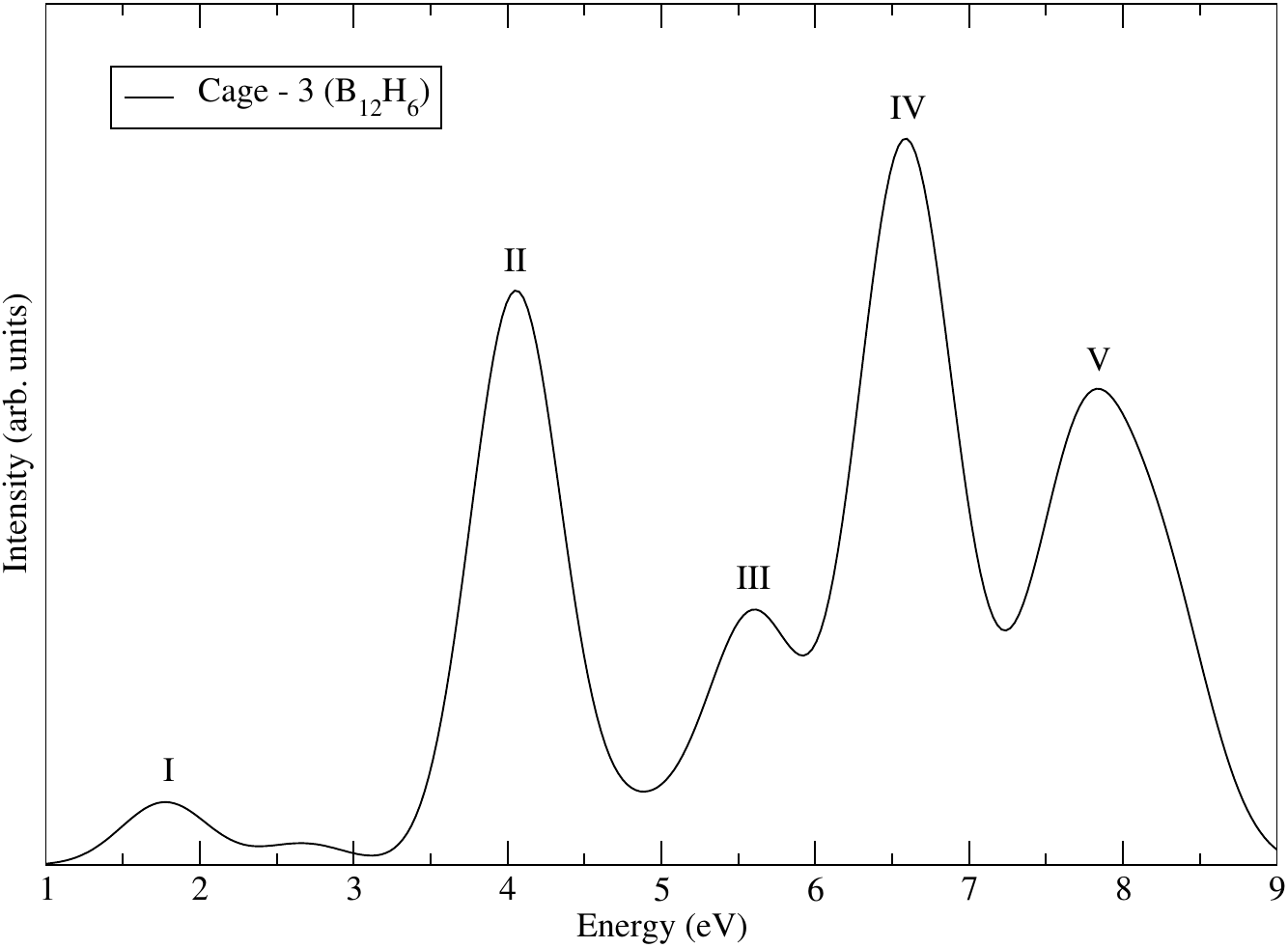}
\par\end{centering}
\caption{The optical absorption spectrum of cage-3 (B$_{12}$H$_{6}$) isomer
calculated using the TDDFT approach at B3LYP/6-31+G{*} level of theory.
\label{fig:optics-cage-3-b12h6}}
\end{figure}

This isomer is 1.33 eV higher in energy as compared to the lowest-energy
cage-1 isomer, with the $C_{2h}$ point group symmetry. The optimized
geometry is presented in Fig. \ref{fig:optimized-geometries-1}(c),
and in that the average bond length between two boron atoms is 1.78
$\text{Å}$, while that between boron and hydrogen atoms is 1.20 $\text{Å}$.
The vibrational frequency analysis predicts that this structure is
also a stable one, and its optical absorption spectrum is presented
in Fig. \ref{fig:optics-cage-3-b12h6}, while the information related
to the peaks is represented in Table \ref{tab:wfs-cage-3-b12h6}.
 
\begin{table}
\caption{The peak locations of the optical absorption spectrum of cage-3 isomer
of the B$_{12}$H$_{6}$ computed using the TDDFT approach, along
with the corresponding oscillator strength (OS), and the wave functions
of the excited states contributing to its peaks. \label{tab:wfs-cage-3-b12h6}}

\centering{}%
\begin{tabular}{|c|c|c|c|}
\hline 
Peak & Energy (eV) & OS & Wave function\tabularnewline
\hline 
\hline 
I & 1.77 & 0.027 & $\arrowvert H-1\rightarrow L\rangle$\tabularnewline
\hline 
 &  &  & \tabularnewline
\hline 
II & 4.00 & 0.172 & $\arrowvert H\rightarrow L+2\rangle$\tabularnewline
\hline 
 & 4.17 & 0.081 & $\arrowvert H\rightarrow L+3\rangle$\tabularnewline
\hline 
 &  &  & \tabularnewline
\hline 
III & 5.61 & 0.090 & $\arrowvert H\rightarrow L+5\rangle$\tabularnewline
\hline 
 &  &  & \tabularnewline
\hline 
IV & 6.42 & 0.081 & $\arrowvert H-11\rightarrow L\rangle$\tabularnewline
\hline 
 & 6.57 & 0.064 & $\arrowvert H\rightarrow L+14\rangle$\tabularnewline
\hline 
 & 6.62 & 0.056 & $\arrowvert H-1\rightarrow L+9\rangle$\tabularnewline
\hline 
 &  &  & \tabularnewline
\hline 
V & 7.67 & 0.062 & $\arrowvert H-1\rightarrow L+17\rangle$\tabularnewline
\hline 
 & 7.79 & 0.052 & $\arrowvert H-1\rightarrow L+18\rangle$\tabularnewline
\hline 
\end{tabular}
\end{table}
The absorption starts with a small peak at 1.77 eV, dominated by the
transition $\arrowvert H-1\rightarrow L\rangle$. It is followed by
an intense peak near 4.10 eV due to the transitions $\arrowvert H\rightarrow L+2\rangle$
and $\arrowvert H\rightarrow L+3\rangle$. The peak-III at 5.61 eV
is due to the state whose wave function consists mainly of the configuration
$\arrowvert H\rightarrow L+5\rangle$. The most intense peak, \emph{i.e.},
peak IV is due to the three excited states dominated by the transitions
$\arrowvert H-11\rightarrow L\rangle$, $\arrowvert H\rightarrow L+14\rangle$,
and $\arrowvert H-1\rightarrow L+9\rangle$. The final peak (V) of
the spectrum is due to two excited states, whose wave functions are
largely composed of the transitions $\arrowvert H-1\rightarrow L+17\rangle$
and $\arrowvert H-1\rightarrow L+18\rangle$. Obviously, peaks IV
and V involve orbitals which are far away from the Fermi level. All
the orbitals of this conformer are non-degenerate.

\subsubsection{Borozene }

By means of first-principles calculations, Szwacki \emph{et} \emph{al}.\cite{Szwacki_et_al}
were the first to propose the existence of a completely planar isomer
of B$_{12}$H$_{6}$, with $D_{3h}$ point group symmetry, which,
in analogy with benzene, they called borozene. We performed geometry
optimization of this isomer at the CCSD/cc-pVDZ level of theory ,
and, in agreement with Szwacki \emph{et} \emph{al}.\cite{Szwacki_et_al},
found the structure to have $D_{3h}$ symmetry (see Fig \ref{fig:optimized-geometries-1}(d)),
with the total energy 1.7 eV higher than that of the cage-1 structure.
The outer-ring is composed of two types of B-B bond lengths, \emph{i.e.},
1.84 $\text{Å}$ and 1.69 $\text{Å}$, while the inner ring has uniform
B-B bond lengths 1.89 $\text{Å}$. The connecting bonds between the
inner ring and the outer ring boron atoms are also of two kinds, \emph{i.e.},
1.67 $\text{Å}$ and 1.68 $\text{Å}$, which differ very slightly
from each other. The uniform B-H bond length in the structure is 1.20
$\text{Å}$. We also carried out the vibrational frequency analysis
using the CCSD approach, and the cc-pVDZ basis set, and found it to
be a stable structure. This isomer has a geometry very similar to
that of the triangular B$_{12}$H$_{12}$ discussed above, except
for the six non-planar hydrogen atoms, and slightly different B-B
and B-H bond lengths. 

In an earlier work from our group, Sahu \emph{et al.}\cite{Sahu_et_al}
reported the structure of borozene, optimized at the INDO-HF level
of theory, to have uniform B-B and B-H bond lengths of 1.65 $\text{Å}$
and 1.18 $\text{Å}$, respectively. They also optimized the structure
at the B3LYP/6-311++g(d) level of theory, and found it to have four
distinct B-B bond lengths 1.63 $\text{Å}$, 1.66 $\text{Å}$, 1.81
$\text{Å}$, and 1.86 $\text{Å}$, in addition to the uniform B-H
bond length 1.18 $\text{Å}$.\cite{Sahu_et_al} Clearly, these values
are in very good agreement with our CCSD-level optimized geometry
parameters. 

\begin{figure}
\begin{centering}
\includegraphics[scale=0.4]{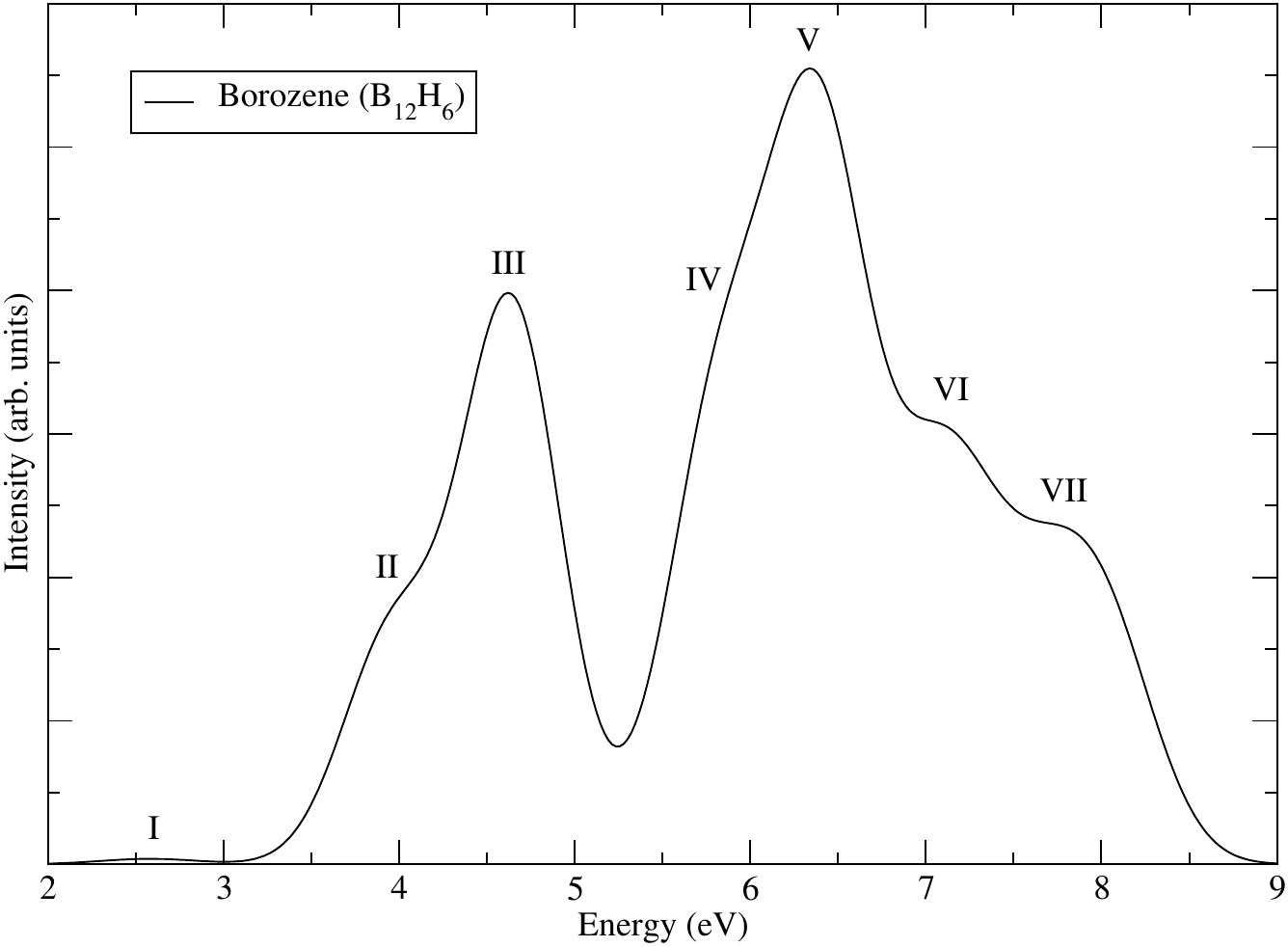}
\par\end{centering}
\caption{The optical absorption spectrum of borozene (B$_{12}$H$_{6}$) isomer
calculated using the TDDFT approach at B3LYP/6-31+G{*} level of theory.
\label{fig:optics-borozene-b12h6}}
\end{figure}

\begin{table}
\caption{The peak locations of the optical absorption spectrum of borozene
isomer of the B$_{12}$H$_{6}$ computed using the TDDFT approach,
along with the corresponding oscillator strength (OS), and the wave
functions of the excited states contributing to its peaks.\label{tab:wfs-borozene-b12h6}}

\centering{}%
\begin{tabular}{|c|c|c|c|}
\hline 
Peak & Energy (eV) & OS & Wave function\tabularnewline
\hline 
\hline 
I & 2.58 & 0.005 & $\arrowvert H\rightarrow L\rangle$\tabularnewline
\hline 
 &  &  & \tabularnewline
\hline 
II & 3.96 & 0.096 & $\arrowvert H-1\rightarrow L\rangle$\tabularnewline
\hline 
 &  &  & \tabularnewline
\hline 
III & 4.64 & 0.240 & $\arrowvert H-1\rightarrow L+1\rangle$\tabularnewline
\hline 
 &  &  & \tabularnewline
\hline 
IV & 5.79 & 0.166 & $\arrowvert H\rightarrow L+3\rangle$\tabularnewline
\hline 
 &  &  & \tabularnewline
\hline 
V & 6.34 & 0.195 & $\arrowvert H\rightarrow L+4\rangle$\tabularnewline
\hline 
 &  &  & \tabularnewline
\hline 
VI & 7.14 & 0.104 & $\arrowvert H\rightarrow L+9\rangle$\tabularnewline
\hline 
 &  &  & \tabularnewline
\hline 
VII & 8.00 & 0.066 & $\arrowvert H\rightarrow L+14\rangle$\tabularnewline
\hline 
\end{tabular}
\end{table}
Our calculated absorption spectrum of this isomer (Fig. \ref{fig:optics-borozene-b12h6})
starts at 2.58 eV, with a weak peak (I), followed by two major peaks
labeled III, and V, along with several shoulders with comparable intensities
which broaden the spectrum. The location of peak I at 2.58 eV is in
excellent agreement with the value 2.6 eV reported by Szwacki \emph{et}
\emph{al}.\cite{Szwacki_et_al}. Because of the strictly planar structure
of borozene, its orbitals can be classified as either of $\pi$ type,
or of $\sigma$ type. On inspection we find that HOMO-1, HOMO, LUMO,
LUMO+1, LUMO+2, and LUMO+3 are all of $\pi$ type, while the immediately
lower/higher orbitals are of $\sigma$ type. As far as orbitals further
away from the Fermi level are concerned, they were found to be of
either type. The information related to various features of the calculated
spectrum of borozene is presented in Table \ref{tab:wfs-borozene-b12h6},
from where it is obvious that the first peak dominated by the HOMO
(H) to LUMO (L) transition is a $\pi$-$\pi^{*}$ excitation. Peak
II at 3.96 eV is due to a state whose wave function consists mainly
of the configuration $\arrowvert H-1\rightarrow L\rangle$, while
the moderately intense peak near 4.6 eV is primarily composed of the
configuration $\arrowvert H-1\rightarrow L+1\rangle$. Clearly, both
these peaks correspond again to\textcolor{red}{{} }$\pi$-$\pi^{*}$
transitions. They are followed by a peak (IV) near 5.80 eV characterized
by the excitation $\arrowvert H\rightarrow L+3\rangle$, which is
also a $\pi$-$\pi^{*}$ transition. The most intense peak at 6.34
eV is identified as a $\pi$-$\sigma^{*}$ transition, due to an excited
state whose wave function is dominated by the single excitation $\arrowvert H\rightarrow L+4\rangle$.
The last two shoulder peaks of the computed absorption spectrum are
located at 7.14 and 8.00 eV, whose wave functions are composed mainly
of excitations $\arrowvert H\rightarrow L+9\rangle$ and $\arrowvert H\rightarrow L+14\rangle$,\textcolor{red}{{}
}which are $\pi$-$\pi^{*}$ and $\pi$-$\sigma^{*}$ transitions,
respectively. In our earlier work\cite{Sahu_et_al}, we also computed
the optical absorption spectrum of borozene using the INDO model,
and the multi-reference-singles-doubles configuration-interaction
(MRSDCI) approach. On comparing our INDO/MRSDCI spectrum\cite{Sahu_et_al}
to the present one, we note that there are very significant differences
between the two. The first peak of the INDO/MRSDCI spectrum was found
at 2.25 eV, as compared to 2.58 eV in the present calculation. Furthermore,
the rest of the peaks in the INDO/MRSDCI spectrum\cite{Sahu_et_al}
were located at energies larger than 7 eV, clearly, in disagreement
with the present calculations. Given the larger basis set involved
in our present calculations, we are inclined to trust these first-principles
calculations more than our previous INDO-model based work.\cite{Sahu_et_al}
However, the disagreement between the two sets of calculations can
definitively be resolved only by experimentally measuring the absorption
spectrum of borozene.

Based on our calculated spectra, we note that the optical signatures
of various isomers of the B$_{12}$H$_{6}$ cluster are sufficiently
different from each other so that they can be identified using optical
spectroscopy.

\section{Conclusions}

\label{sec:conclusions}

With the aim of understanding the stability against distortions, we
performed electron-correlated calculations on various conformers of
B$_{12}$H$_{12}$ and  B$_{12}$H$_{6}$ clusters. The geometries
of all the isomers were optimized using the CCSD approach and cc-pVDZ
basis set. In order to investigate the structural stability, the vibrational
frequency analysis was also carried out for all the isomers . According
to our calculations, all the four conformers of B$_{12}$H$_{6}$
considered in this work are stable, while only the deformed-chain
isomer of B$_{12}$H$_{12}$ is stable. In particular, the perfect
icosahedral cage of B$_{12}$H$_{12}$ ($I_{h}$ symmetry) cluster
was found to be unstable, consistent with the Jahn-Teller analysis.
The same analysis suggests that the di-anion B$_{12}$H$_{12}$, with
the perfect $I_{h}$ symmetry should be stable, in full agreement
with our calculations. We also computed the optical absorption spectra
of all the stable isomers using the TDDFT approach, and found that
they differ significantly from each other, suggesting the possibility
of their identification using optical spectroscopy.

\section*{Author Information }

\subsection*{Corresponding Authors}

Alok Shukla:  {*}E-mail: shukla@phy.iitb.ac.in

\subsection*{Notes}

The authors declare no competing financial interests.

\section*{Acknowledgements}

Work of P.B. was supported by a Senior Research Fellowship offered
by University Grants Commission, India.

\section*{TOC Graphic}

\begin{figure}[H]
\includegraphics[width=8.25cm,height=4.45cm]{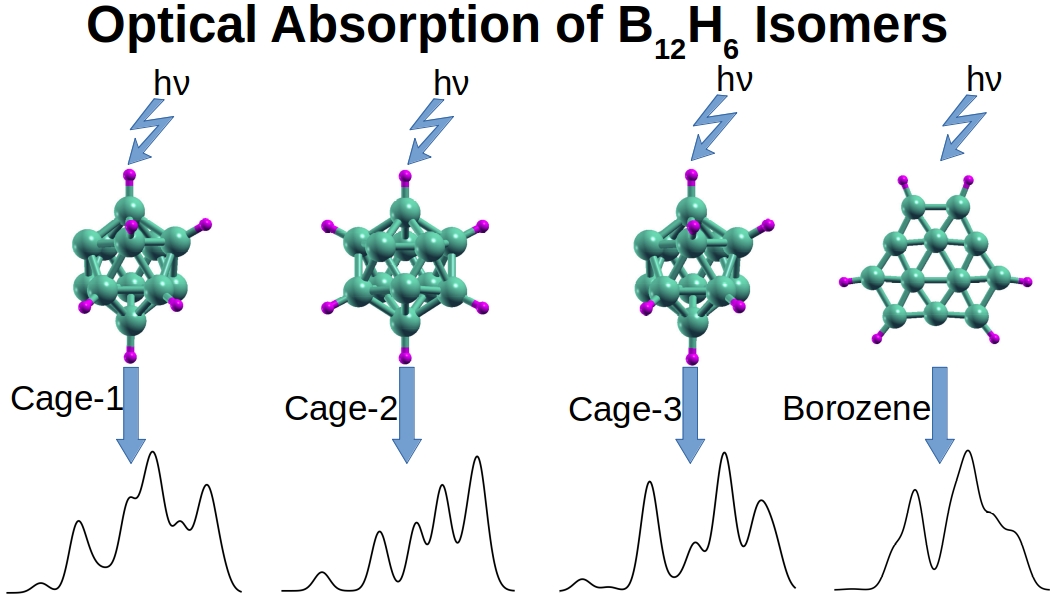}
\end{figure}

\addcontentsline{toc}{section}{\refname}\bibliography{boron_hydrides}

\end{document}